\documentclass[a4paper,%
   a4paper,%
   BCOR12mm,%
   11pt,%
   abstracton,%
   pointednumbers,%
   tablecaptionabove,%
   footinclude,%
   halfparskip,%
   ]{scrartcl}
\usepackage[ilines]{scrpage2} 
\usepackage{epsfig}
\usepackage{multicol}
\usepackage{amsmath}
\usepackage[round,colon,sort,authoryear]{natbib}

\newcommand{\UPLB}{University of the Philippines Los Ba\~{n}os}

\title{\Large A System for Sensing Human Sentiments to\\Augment a Model for Predicting Rare Lake Events}
\author{\large Jaderick P. Pabico\\
   \normalsize{Research Collaboratory for Advanced Intelligent Systems}\\
   \normalsize{Institute of Computer Science, \UPLB}\\
   \normalsize{College 4031, Laguna, Philippines}}
\date{}

\lehead{}
\rohead{\small\thepage}
\lofoot{\small jppabico/uplb/a system for sensing human sentiments...}
\cofoot{}
\rofoot{}

\begin{document}
\maketitle

\begin{abstract}
Fish kill events (FKE) in the caldera lake of Taal occur rarely (only 0.5\% in the last 10 years) but each event has a long-term effect on the environmental health of the lake ecosystem, as well as a devastating effect on the financial and emotional aspects of the residents whose livelihood rely on aquaculture farming. Predicting with high accuracy when within seven days and where on the vast expanse of the lake will FKEs strike will be a very important early warning tool for the lake's aquaculture industry. Mathematical models to predict the occurrences of FKEs developed by several studies done in the past use as predictors the physico-chemical characteristics of the lake water, as well as the meteorological parameters above it. Some of the models, however, did not provide acceptable predictive accuracy and enough early warning because they were developed with unbalanced binary data set, i.e., characterized by dense negative examples (no FKE) and highly sparse positive examples (with FKE). Other models require setting up an expensive sensor network to measure the water parameters not only at the surface but also at several depths. Presented in this paper is a system for capturing, measuring, and visualizing the contextual sentiment polarity (CSP) of dated and geolocated social media microposts of residents within 10km radius of the Taal Volcano crater ($14^\circ$N, $121^\circ$E). High frequency negative CSP co-occur with FKE for two occasions making human expressions a viable non-physical sensors for impending FKE to augment existing mathematical models.
\end{abstract}


\section{Introduction}
Fish kill events (FKEs) in the caldera lake of Taal in the province of Batangas occur infrequently with recorded daily frequency of only 0.5\% for the last 10 years. When FKEs occur, however,  their consequences to the lives of the people whose livelihood rely on fish farming is devastating. Sensing  with higher accuracy, therefore, when in the next seven days and where in the vast expanse of Taal Lake will FKE occur will be an important early warning tool for the Taal Lake aquaculture industry. The development of such a high fidelity sensing system will not only be interesting to the practioners of the fields of limnology, but also to those who are in the practice of fisheries science, meteorology, economics, predictive analytics, and agricultural and biosystems engineering. 

The prediction of the occurrences of fish kill events (FKE) in Taal Lake is challenging to practitioners in these fields, not only because of its trivial impact to the livelihood of the human population who rely on the lake, but also because the lake is a very unique subject of study compared to other aquaculture lakes in the world. Geographically, the lake was formed over a caldera and is host to Taal Volcano, a famous tourist attraction known to be the smallest active volcano in the world, which stands on a smaller lake within an island, which in turn is located in the middle of Taal Lake. 

FKE in Taal Lake may not only be the result of the already known usual chemico-physical factors affecting most lakes worldwide, but may also be worsen by the presence of volcanic vents that actively spew sulfuric emissions. The dynamics of these emissions do not only result to having extremely high temperature difference between the surface and bottom water, but additionally the jets of high-temperature streams will cause mechanical stirrings of the lake bottom. As of this writing, no studies have been conducted that can quantitatively predict when such emissions occur and at what magnitude, much so where these thermal vents are spatially located on the lake floor despite the few attempts to map the bottom with acceptable-resolution bathymetry. Even if high-resolution bathymetry from high-fidelity sonar technologies can be obtained for Taal Lake, because Taal Volcano is an active volcano, new volcanic vents form while the present ones undergo an unpredictable cycle of closing down and opening up depending on various tectonic activities that occur (mostly are hypothesized to be undetected) within the Luzon geological plate.

Despite of these challenges, several studies have been made in the past to predict the Taal Lake FKEs through predictive modeling using as predictors the physico-chemical characteristics of the lake water in combination with the lake's immediate meteorological parameters~\citep{Macandog12a,Macandog12b,Macandog12c,Macandog13a,Macandog13b}. Most of these models, specifically those that used mined patterns from historical data, faced the problem of unbalanced binary dataset, i.e., having dense negative daily data (no FKE) over sparse positive daily data (with FKE). Note that the granularity of temporal variables in these dataset was set to daily basis because this is the finest granularity that meteorological data can be obtained from Philippine Atmospheric, Geophysical and Astronomical Services Administration (PAGASA) Station in Ambulong, Tanauan City, Batangas ($14.083^\circ$N, $121.050^\circ$E, 10.0MASL), the nearest and only meteorological station in the whole province. Although the physico-chemical characteristics of the lake water at various depths as recorded and curated by the Bureau of Fisheries and Aquatic Resources in Region IV-A (BFAR4A) is in the daily granularity, the frequency when the reading of these dataset does not happen on a regular daily basis. Most critical characteristics are only measured when there is a reported observation of the symptoms of FKE. Thus, the physico-chemical dataset contains a lot of missing records. Even though standard statistical regression and time-series techniques will work for datasets with missing records, they are not appropriate for modeling patterns characterized by dense-negative and sparse-positive binary records. 

Probabilistic methods, such as the Bayesian network of models inferred from historical qualitative and quantitative unbalanced dataset provides an acceptable predictive rate  and accuracy but is not realistically useful in Taal because the models require a vast network of physico-chemical sensors. Before the Bayesian network of models can be fully utilized, the government in cooperation with the private sector must invest highly in the development, operation, and maintenance of high resolution sensing infrastructure that will cover not only the entire surface of the lake but also that which will expand several meters deep along the lake depth, preferably at least five meters more than the average depth of fish cages. If the Bayesian network of models prove to be accurately useful when this network of sensors is put in place, then such an investment is projected to be costlier than the savings the fish industry will incur in avoiding FKEs. Thus, probabilistic modeling methods that rely on the input from expensive sensor readings are not an economically worthy undertaking.

While the models developed using probabilistic means may require high investment for the government and stakeholders, does it mean that these kind of undertakings, however accurate and useful the results are, will be put to waste? Definitely not if a new kind of sensors can be deployed at the lake without spending much for its development, operation, and maintenance. In recent years,  non-physico-chemical sensors have emerged as a new ``device'' to somehow sense the environment with high-fidelity and utmost recency. These ``devices'' are called ``social sensors.'' Social sensors measure the apparent general sentiment of the human population where environmental, social, political, or economic events such as earthquakes, storms, disease outbreaks, population disturbances, sporting events, movie premiers, political campaigns, and stock market crashes are currently happening. The population's general sentiment can easily be read and then measured by processing the online postings of users on the social media over the Internet. The sentiment can be computed using a network of computers as soon as the user posts an experience, idea, or opinion. The utility of this emerging sensor has already been exploited in the works of~\citet{Sakaki10, Aramaki11, Fraustino12, Lampos12}, and~\citet{Li13}, where negative (or positive) sentiments co-occur at the same general location (and direction) of an environmental disaster (or happy social event).

In a past effort~\citep{Pabico14b}, the user-generated microblogs from the social media Twitter were studied to find out whether a high frequency of words that describe FKE in Taal Lake co-occur with the actual fish kills. It was found out that words such as ``maitim na tubig'' (black water), ``amoy asupre'' (sulfuric odor), and ``nahibay na isda'' (dizzy fish) among others, have been observed from microblogs originating from within the 10Km radius of the peak of Taal volcano crater ($14^\circ$N, $121^\circ$E). Geo-locating microblogs are not difficult to do because most of them are geotagged and time-stamped, especially those that were sent through GPS-equipped smart phones. Most fish cages in Taal Lake, as well as the communities whose livelihood rely on fish farming, are located within this radius. Through years of experience, residents in these communities have already compiled observable symptoms (COS) for FKE which they use as a form of adaptive mechanism to prepare them for FKE and to warn others of the onset of FKE in their locality. It is very interesting to note that the increase in the observed frequency of words in the COS co-occur with the recorded FKEs, as well as when the lake water quality level is deemed  critical  by BFAR4A, suggesting that the frequency may be a good indicator for an onset of FKE. However, FKE are so locally-specific that they may strike one fish cage, but nearby fish cages (even as near as 5m) will not be affected. Further, COS from one location may not be the same with the others. For example, there are fish cages which are more often affected by ``amoy asupre,'' suggesting thermal vents at the nearby (though not directly) bottom, than by ``berdeng tubig'' (green water). Thus, the COS in these locations does not include ``berdeng tubig.'' On the other hand, there are fish cages which are always affected by ``berdeng tubig'' than by ``amoy asupre.'' Intuitively, the COS in these areas does not include ``amoy asupre.'' In either case, ``nahibay na isda'' were observed, suggesting dissolved oxygen depletion. It has already become a local knowledge by the residents in the area that FKE always follow after ``nahibay na isda'' is observed for an extended period of time.

The general Twitter microposts of residents in the Taal Lake communities may not include the COS, but this does not necessarily mean that the poster does not experience negative feelings (e.g., anxiety) over the possibility of a FKE in their area. In fact, a huge percentage of polled microposts from the Taal Lake area during the onset of FKE do not contain COS. However, when the sentiments of the microposts were measured, a huge percentage of the polled samples was found out to be negative, suggesting that anxiety, sadness, fear, and probably depression are being experienced by the members of the polled population. Specifically, an increasing frequency of negative contextual sentiment polarity (CSP) of microposts was observed to have co-occurred 1-6 days prior to a FKE during the reported events in February 2013 and in January 2014. During times when symptoms of FKE are not reported, the general sentiments of microposts range from neutral to positive (e.g., optimistic, happy, and elated). This suggests that there does exist evidence that FKE's in Taal Lake and negative CSP in microposts of Taal Lake residents co-occur. If the frequency of negative CSP increases, one may be able to prepare for an FKE to occur within six days.

\section{Review of Literature}
\subsection{The Social Media}

\begin{figure}[htb]
\centering\epsfig{file=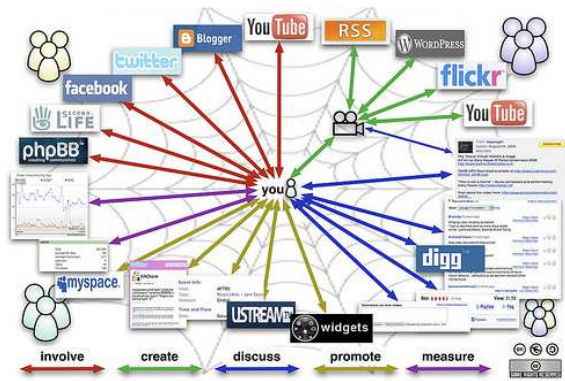, width=3in}
\caption{A popular visualization of the conceptual framework of the Social Media by~\citet{Hayes08} out of the many visualizations that exist. Here, the social media is visualized as a social web.}\label{fig:1}
\end{figure}

The advent of the so-called social media over the Internet has impacted the way people live in the digital age. The social media has become a ubiquitous tool for people to meet, communicate, and collaborate with other people. Some examples of social media sites are Google+, Facebook, Digg, Flickr, YouTube, and Instagram~\citep{Page98, Zuckerberg04, Adelson04, Butterfield04, Chen05, Systrom10}. Figure~\ref{fig:1} shows one of the many visualizations of the inferred conceptual framework of the current social media as utilized by and in the point of view of an account owner~\citep{Hayes08}. 

The exponential proliferation of social media sites offering varied services and tools for meeting, communicating, and collaborating with other people has significantly transformed how the so-called connected people's lives. Now, the social media sites are expected to function as real-time, on-site sources of trusted information, as well as a medium for reporting events as they happen. For example, in the recent natural disasters that hit the Philippines, particularly the 2013 Bohol Earthquake, the strongest tropical cyclone ever recorded that made landfall Haiyan (Typhoon Yolanda), and the tropical cyclone Rammasun (Typhoon Glenda)~\citep{Fischetti13, Mangosing13, Marquez13, Masters13,   Larano14}, people in the social web have extensively used the services either as sources of breaking news or as means for sharing information that they have experienced first hand. The services can be used in real-time that users were given the capability by the social web to share their experiences as the event happens, often validated by different people who also happen to experience the same~\citep{Nagar12}. This seemingly real-time and interactive capabilities were not quite possible using the technologies used in broadcast communication, where information only flows from the broadcaster to the target audience.

The trust that people give to the social media as a source of real-time and reliable information has already equaled, if not surpassed, the trust that they give to usual authoritative sources of real-time information like the television and the broadcast radio. In fact, adults in the U.S. have reported that the Internet was their preferred source for information and they ranked it as the most reliable source for news~\citep{Zogby09}. The public's online activities increase during disasters because the people have increasingly turned to the social media for the most up-to-date information. For example, \citet{Nagar12} have reported that the social media site Twitter was the most active site during the 2011 Japanese tsunami where users worldwide posted more than 5,500 tweets per second just after the disaster struck. Additionally, it was also observed that there is an increase in the presence of non-journalists who practice news reporting during natural calamities~\citep{Caragea14}. These are average citizens present in the location of the calamity and report happenings through the social media. Not only the social media is used for seeking and sharing information during disasters but, in addition, users expect emergency managers to respond to victims through real-time monitoring the social media posts.  This fact is echoed by 75\% of the 1,078 respondents surveyed by the American Red Cross in 2010 who said that they ``expected help to arrive within an hour if they posted a request on a social media site.''

\subsection{Microblogs}

One of the most important attributes of the social media that in recent years has increasingly been relied to by the people at large is its capability to allow users to post any information in real-time. Although the social media has services that allow sharing of events in visual forms through posting of amateur video clips (for example in YouTube), of photographs taken from hand-held cameras (e.g., in Instagram), or of computer documents (e.g., in Google+), ``any'' information also includes personal experiences of the users that are often expressed as written texts. The social media site Twitter~\citep{Dorsey06} is one of the services in the Internet that allows posting of short texts that are often updated by users several times in a day. The posts, each containing up to 144 characters and aptly called a microblog or tweet, often reflect the uninhibited spurt-of-the-moment sentiment of a user to verbalize in written form a current personal experience. Aside from tweets, Twitter allows the sharing of micromedia like photographs, video clips, and audio clips~\citep{Sakaki10}.

Because of the uninhibited nature of the microblogs, other users who particularly have close ties to a user~$\mathcal{U}$ often come back to the social media site several times in a day to check what~$\mathcal{U}$ is doing and what~$\mathcal{U}$ is thinking about now. Oftentimes, the user~$\mathcal{U}$ is a celebrity like the internationally renowed singer Katie Perry,  US President Barack Obama, and the Roman Catholic's Holy See Pope Francis, whose number of users who follow them are in the range of millions. In fact, as of early January this year, they respectively have 49.0M, 40.7M, and 3.0M followers~\citep{Pabico14}, but recent report ten months after~\citep{vanZanten14} shows that they already have 60.7M, 50.5M, and 4.8M followers. For some personalities in the show, entertainment, and advertising businesses, the number of Twitter followers correlates to success in product endorsements, which necessitates the daily frequent update of the microblogs to keep the followers, if not to maintain a positive growth in their number.

Second only to personal sentiments, a large number of microblog updates results in numerous reports related to social events, such as parties, sports, and political campaigns. Some events  include disasters such as storms, ﬁres, traffic jams, riots, enundations from heavy rainfall, and earthquakes. Actually, it has been documented by researchers that Twitter has been used for various real-time disaster response mobilization activities especially for users seeking help during a large-scale ﬁre emergency and live traffic updates. Adam Ostrow, the Editor in Chief of a social media news blog Mashable, opined the following about the interesting phenomenon of the real-time media:

\begin{quote}
Earthquakes are one thing you can bet on being covered on Twitter ﬁrst, because, quite frankly, if the ground is shaking, you’re going to tweet about it before it even registers with the USGS and long before it gets reported by the media.
\end{quote} 

This report has been the motivation behind several independent studies which used  a new method called crowdsourcing to ``sense'' the environment not by any physico-chemical devices but through what is called social sensors. Crowdsourcing is the process of obtaining sensed data from the seemingly numerable independent and unrelated humans, while social sensors are humans who are currently experiencing a disaster.  Examples of studies that reported success in using crowdsourced social sensors are identifying the rate of contagion in flu epidemics~\citep{Aramaki11, Li13}, predicting the epicenter of earthquakes~\citep{Sakaki10}, and nowcasting the trajectories of storms~\citep{Fraustino12}. Nowcasting, a contraction of now and forecasting, is the forecasting of observable physical phenomenon, such as the weather and the economy, within the next six hours with reasonable  accuracy~\citep{Lampos12}.
 
With the growing number of people residing near different disaster-prone areas, the need for a tailor-fit, updated, and fast warning and response system is increasing~\citep{Mandel12, Caragea14} and Twitter is one of the tools that are being used for this purpose. Its real-time nature makes it a good outlet of messages for reporting and sharing disaster-related news, and even for asking help~\citep{Li13, Caragea14, Sakaki10}. Because of its wide-spread utility, not only news agencies use the Twitter social media to broadcast news, but also science researchers who analyse the population sentiment during disasters using the Twitter's extensive data stream as input~\citep{Pak10}.

\subsection{Sentiment Analysis}

\em{Sentiment Analysis} (SA), also called \em{opinion mining}, is the use of technologies developed in the Computer Science's subfield of Natural Language Processing (NLP) to automatically analyze (usually written) texts and identify the prevailing sentiments in these texts. Sentiments identified are attitudes, emotions, or opinions towards a certain observed phenomenon. Recently, SA has been utilized in a number of disciplines like business, finance, and politics. In finance, it is used to measure the investment mood of people~\citep{Sanchez-Rada14}, while in politics, it is used to determine the pulse of the voters and predict a possible winner of an election~\citep{Bravo-Marquez13}.

There are two methods commonly used to categorize sentiments for SA: machine learning, and lexicon-based. Machine learning (ML) is the use of a combination of various proper computational intelligence techniques that allow a computer to learn patterns from examples~\citep{Kovahi98, Kruse13}. The computational patterns to learn in this aspect is to classify the contextual polarity of texts into either positive or negative sentiments~\citep{Shalunts14}. Sometimes, a neutral category is added~\citep{Koppel06}. Examples of computational techniques that are commonly used for SA are Naive-Bayes~\citep{Gamallo14}, Support Vector Machines~\citep{Mullen04}, and Artificial Neural Networks~\citep{Sharma12}. On the other hand, the lexicon-based method uses a set of words called sentiment dictionary or lexicon, wherein each word is associated to either positive, neutral, or negative sentiment~\citep{Taboada11}. When a word~$w$ is encountered, the sentiment dictionary is used as a look-up table for assigning CSP to~$w$~\citep{Palanisamy13}. This makes the lexicon-based method easier to implement than the machine learning ones. It is not difficult to combine the two methods to come up with a better SA as evident in the work of~\citet{Zhang11} as both methods are not mutually exclusive.

Formally, SA is a transformation~$T$, such that $T(w) \rightarrow s \in \{–, 0, +\}$, where~$T(w)$ is the transformation of a word~$w$ to a CSP~$s$, which is either negative ($–$), neutral ($0$), or positive ($+$). Note here that~$T$ is a transformation function that could either be generated by any ML heuristic, a look-up table based on a lexicon of words, or a combination of both. The transformation~$T$ is oftentimes the core function of an automated classifier~$C$ which accepts as inputs a sentence or phrase~$I$, and outputs~$O$ as the sentiment of~$I$. Converting from~$I$ to~$O$ includes the following steps: 
\begin{enumerate}
\item Partition~$I$ into parts of a sentence or a set of words $W=\{w_1, w_2, \dots, w_n\}$ using a NLP technique~\citep{Liu10, Groh11};
\item Use~$T$ on each $w_i \in W$, $\forall i = 1, 2, \dots, n$, to obtain a set of CSP $S = \{s_i | i = 1, 2, \dots, n\}$;
\item Provide a final CSP score over~$S$ using some function, the simplest of which is an averaging one: $S_\mathrm{mean} = n^{-1} \times \sum_{i=1}^n s_i$; and
\item Return $S_\mathrm{mean}$ as~$O$.
\end{enumerate}

\subsection{The Twitter Tweet API and Its Utility}

The Twitter Tweet Application Programming Interface (TAPI) is a set of computer commands provided by the Twitter developers for exclusive use of programmers to allow them to tap into the Twitter data stream and gather tweets at a specific timeframe and geo-location~\citep{Dorsey06}. Nowadays, most tweets that are sent from mobile smart phones are already geo-tagged because of the device's capability to receive geo-positional data from Global Positioning System (GPS) satellites. Once the streamed tweets have been collected by a computer program that uses the TAPI, they will become inputs~$I$ to an automated classifier~$C$ which will output the CSP~$O$ of the tweets. 

With the timestamp~$t$, geolocation $g = (x, y)$ where~$x$ and~$y$ are respectively longitude and latitude data, and CSP~$O$, a tweet data can be formalized as a triple $(t, g, O)$. A collection of these triples over an area within some time interval can be plotted in a time-evolving Geographic Information System (GIS) to visualize the dynamics of a population's sentiment. This idea resulted in various applications for the TAPI in near real-time visualizing and tracking earthquakes, hurricanes, typhoons,  diseases, and other natural phenomena~\citep{Sakaki10, Aramaki11,  Fraustino12, Lampos12, Mandel12, Li13}.

\subsection{Real-time Tracking of Natural Disasters and Diseases}

A number of computer applications using the TAPI have been devised to track natural disasters and diseases in near real-time. The earliest reported application is that of~\citet{Sakaki10} where the created computer system monitors the tweets of Japanese users and detects from the tweets mentions of experiences of earthquakes with high probability. The system notifies Twitter users in Japan much faster than the broadcast warnings of the Japan Meteorological Agency (JMA). The basic idea of the application is that each Twitter user acts as a ``social sensor'' which (or who) detects an event (i.e., an earthquake) and reports the experience with certain probability. Since a tweet includes the timestamp~$t$ and the geolocation data~$g$, the computer system can then infer the trajectory of the shock propagation of an earthquake over an area and can send early warning to the potentially affected population.

A similar system, where not only mentions of earthquake events were detected but also the CSP's of online posts in both Twitter and Facebook, was subsequently created by~\citet{Doan12} and then further improved by~\citet{Vo13}. In the work of~\citet{Doan12}, 1.5~million online posts were investigated from 9 March 2011 to 31 May 2011 to track the awareness and anxiety levels of the residents in the Tokyo Metropolitan District in relation to the 2011 Tohoku Earthquake and the tsunami and the nuclear emergencies that follow. An improved system was used in the work of~\citet{Vo13} where they also detected other emotions (e.g., unconcerned, concerned, calm, unpleasant, sad, fear, and relief) in addition to the anxiety of the population affected by the Japan earthquakes of  11 March, 7 April, 4 April, and 10 July all in 2011. The studies found out that SA in online posts relating to a sequence of disasters (earthquake, tsunami, and nuclear emergency) is a good early warning system for the target population and a useful resource for tracking the dynamics of the population's general sentiment. The studies emphasized the resiliency of the Japanese people in facing a series of disasters as the SA showed the anxiety and fear levels of the population quickly returning to normal within the day after the disasters.

Starting from the reported system of~\citet{Sakaki10}, several other researchers followed in  tracking the sentiment of a population that are affected by natural disasters using a combination of the Twitter API and SA such as those of Hurricane Irene in August 2011~\citep{Mandel12}, of Hurricane Sandy in October 2012~\citep{Caragea14}, of the flooding in the Philippines caused by the Southwest monsoon (or Habagat) in August 2012~\citep{Lee13}, and of the flooding in Germany and Austria in June 2013~\citep{Shalunts14}.  Not only natural disasters can be tracked but also man-made crises such as the gaspipe explosion of September 2010 in San Bruno, California~\citep{Nagy12}. Similar systems were also created to track disease epidemics such as the surveillance of influenza-like illnesses in several regions of the United Kingdom~\citep{Lampos10}, of the pandemic outbreak of H1N1 in 2009~\citep{Chew10}, and of several cases of Dengue epidemics in Brazil~\citep{Gomide11}. These studies found that tweets about a certain event (disaster or epidemic) originate from different locations, and during the peak of the event, the CSP of the tweets are at its most negative and concentrated within the location where the event strikes. After the peak of the event, CSP's of tweets connected to the disaster vary at different locations~\citep{Caragea14}. \citet{Nagar12} observed that the tweet sentiments clutter, while~\citet{Vo13} noticed that the CSP's went back up to normal within 24 hours. Both studies agree, however, that the strength of CSP's are mostly based on the tweets' distance from the event. Most researchers opined that these CSP's are helpful in building social awareness of the phenomenon.

\subsection{Twitter and Taal Lake FKE}

Building from the various works mentioned above and the perceived significant utility of the Twitter micropost as ``social sensors'' for FKE in Taal Lake, a preliminary study was conducted to test whether certain tweet features will co-occur with a reported fishkill~\citep{Pabico14b}. Words that are included in the COS, such as amoy asupre, berdeng tubig, maitim na tubig, and nahibay na isda were observed to have gained high frequency of occurrences in tweets whose geo-location is within the 10Km radius of the peak of Taal Volcano ($14^\circ$N, $121^\circ$E). The frequency of occurrences increased above the normal within one week before two officially reported FKEs in February 2013 and in January 2014. These suggest that the Twitter users from within the Taal Lake area and whose livelihood depend on fish farming are talking (and probably reporting) to each other through Twitter about the observed symptoms of fish kills. The more the fishfolks talk about it means that they have observed the symptoms much more frequently. One can already infer that an onset of an FKE is already happening.

This current work is built on the earlier work of~\citep{Pabico14b} and improves it with the computation of the CSP's of tweets. The earlier work provided a basis for considering the frequency of words in the COS as an indicator for FKE. However, not all tweets in the area include words in the COS. Most of the tweets contain quantifiable sentiments that could provide with high certainty that as the polarity of the tweets become more negative, one can measure that the residents are becoming anxious as a result of their direct observation and  perceived water quality of the lake surface.

\section{Methodology}
\subsection{Programming a Scraper with TAPI}

A computer program termed here as {\tt pScraper} was written to automatically scrape the Twitter social networking site for tweets originating from within the 10Km radius of the Taal Volcano crater (at geo-location $14^\circ$N, $121^\circ$E). Scraping is the process of extracting pertinent data from web pages obtained from crawling the Internet. Crawling a set $W_n$ of $n$ web pages $W_n=\{p_0, p_1, \dots, p_{n-1}\}$ means downloading the subset $W_{n-1} = \{p_1, p_2, \dots, p_{n-1}\} \subset W_n$ web pages given the initial web page $p_0$. From the respective uniform resource locator (URL) links in hypertext markup language (HTML) anchor tags found in a web page $p_i$, the next web page $p_j$ can be obtained and whose respective data can be scraped, $\forall p_i,p_j\in W_n$, $i\ne j$.
 
Twitter provided TAPI to allow for the automatic scraping of Twitter pages~\citep{Dorsey06}. The automatic scraping of tweets originating from within 10Km radius of $14^\circ$N, $121^\circ$E was performed from 03 February 2013 to 30 January 2014. The Twitter posts collected are from 27 January 2013 to 30 January 2014 (369 days) where only two FKE were recorded, one that happened in 02 February 2013 and another in 16 January 2014. 

TAPI allows for the collection of Twitter posts up to seven days earlier~\citep{Dorsey06} and so posts before the 03 February 2013 FKE were obtained during the start of the scraping process. The code for {\tt pScraper} was written in Perl V5.10.1. 

\subsection{Collecting and Archiving Tweets}

Using {\tt pScraper}, a total of a little over 3.4M different tweets were collected. These tweets were sent by 62,569 unique Twitter account users during the scraping period from 03 February 2013 to 30 January 2014. A microcomputer platform with a 2-core 2GHz 686-based processor, 1GB shared primary memory, 80GB secondary memory, 100mbps Ethernet-based network interface, and Version 6.4 of Scientific Linux Distribution of 64-bit Gnu/Linux Operating System with Gnome 3.8.4 XWindows Manager was used to run {\tt pScraper}. Since tweets can be sent by the account users anytime, {\tt pScraper} was scheduled to automatically run by the standard Gnu/Linux scheduler Cronie V1.4.4 every 15min for a total of 35,424 runs (i.e., 369 days $\times$ 24 hrs/day $\times$ 4 runs/hr) during the scraping period. 
Because of the enormous amount of collected data, all tweets including their metadata such as the timestamp~$t$, geolocation~$g$, and Twitter account user~$\mathcal{U}$ were saved in a relational database system (RDBMS) for a much more efficient data archiving and processing system. The RDBMS used to archive all 2.4M tweets including each tweet's metadata is MySQL V5.1.69. 

\subsection{Computing the Sentiment Polarity of Tweets}

A classifier (called here as {\tt pClass}) using a lexicon-based sentiment analyzer with the English lexicon coming from the works of~\citet{Taboada11} was created. Filipino words and their corresponding sentiments were added manually to the lexicon. Each of the tweets in the database was then fed as input to {\tt pClass} while {\tt pClass}' subsequent output was then saved with the same record as the input into the RDBMS. Similar to {\tt pScraper}, the code of {\tt pClass} was also written in Perl V5.10.1. The program {\tt pClass} was run every midnight during the scraping period for a total of 369 runs. Thus, just right after every midnight of any day $d_i$ during the scraping period, the CSP of the collected new tweets during day $d_{i-1}$ were computed, and that the CSP of all newer tweets that were added during day $d_i$ were computed several minutes at the start of day $d_{i+1}$.

\subsection{Temporal Analysis of CSP}

The temporal variation of CSP was conducted using several time frames: hourly, daily, weekly, and monthly. Since the timestamp of each tweet with its CSP is saved in the RDBMS, the data for each of these time frames were extracted from the respective simple SQL queries to the RDBMS. The hourly analyses (hourly total and average) were conducted to see if there is diurnal variability in the sentiment and to capture time-of-day effects in the sentiment of the Twitter account owners. The daily, weekly, and monthly analyses (also total and average) were also conducted to see seasonal variabilities in the sentiment at various granularities (daily, weekly, and monthly granularity, respectively). The respective data were input to Scientific Linux's standard data visualization program GnuPlot V4.2.6 that resulted to the respective temporal plots of various time frames. This paper, however, only presents the daily analysis due to space constraints. The respective analyses for the hourly, weekly, and monthly time frames will be presented in the archival publication version of this paper.

The FKE data during the scraping period was superimposed to the various temporal plots to visually see if a pattern in CSP, for any time frame, will have an observable co-occurrence with the FKE. Only visual analysis was performed and was assumed to have sufficed because the number of FKE record is very small, i.e., two occurrences: one in 02 February 2013 and another in 16 January 2014. A more sophisticated analysis involving machine learning techniques for mining big data and conducting analytics will be used in the future once enough FKE data with the same time range coinciding with the tweet collection is available.

\section{Results and Discussion}

\subsection{The Scraper}

\begin{figure}[htb]
\centering\epsfig{file=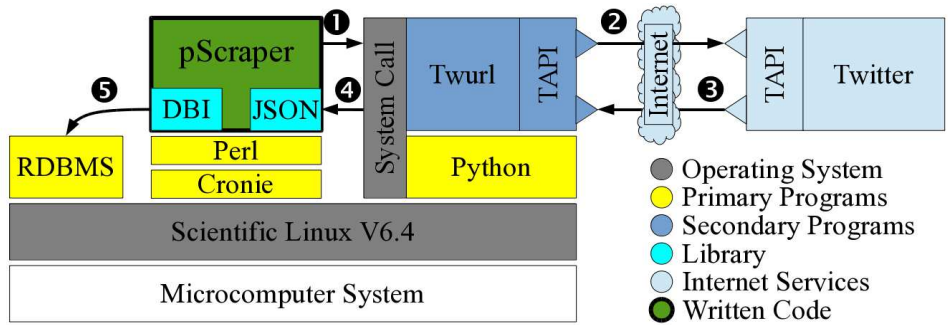, width=5in}
\caption{The conceptual relationship and data exchange among pScraper, Perl, Twurl, Python, JSON, and TAPI: (1)~{\tt pScraper} calls {\tt Twurl} via a system call; (2)~{\tt Twurl} uses TAPI to connect to Twitter via various Internet communication protocols; (3)~Twitter uses TAPI to return a JSON-formatted tweet data through various Internet protocols; (4)~{\tt Twurl} returns to {\tt pScraper} the JSON-formatted tweet data; and (5)~{\tt pScraper} sends the tweet data to the RDBMS for archiving through DBI.}\label{fig:3}
\end{figure}

To be able for the {\tt pScraper} to collect data seamlessly from Twitter using TAPI, it calls {\tt Twurl} via Scientific Linux's system call. {\tt Twurl} is a Python-based program that connects to Twitter through various Internet data communication protocols and uses TAPI to obtain tweet data. {\tt Twurl} returns to {\tt pScraper} a JSON-formatted response from Twitter which {\tt pScraper} reads using Perl's JSON library. The collected tweets are then archived by {\tt pScraper} to the RDBMS via Perl's DBI library. Figure~\ref{fig:3} shows the conceptual relationship and data exchange among the computer programs {\tt pScraper}, Perl, {\tt Twurl}, Python, JSON, and TAPI. The figure represents two conceptual relationships (vertical and horizontal) between any two machines (actual or virtual). Vertical relationship means that the machine represented by the block at the bottom is running a machine that is directly above it, while horizontal relationship between any two machines means data exchange is possible between them. For example, the microcomputer system runs Scientific Linux, which in turn runs Cronie, which in turn runs Perl, which in turn runs {\tt pScraper}, while {\tt pScraper} may exchange data with RDBMS.

With {\tt pScraper}, the scraping process does not only result to gathering the raw data from the tweeter posts but also to identifying the relationships between any two pages $p_i$ and $p_j$, $p_i,p_j\in W_n$, $\forall i\ne j$. The identification of relationships between any two pages though their respective HTML anchors allows for the inference of the topology of the network $\mathcal{N}(W_n,L)$ of web pages. Thus, from $W_n$, the set of links $L = \{(i, j)| p_i, p_j\in W_n\}$ can be obtained. 

\subsection{The Collected Tweets}

\begin{figure}[htb]
\centering\epsfig{file=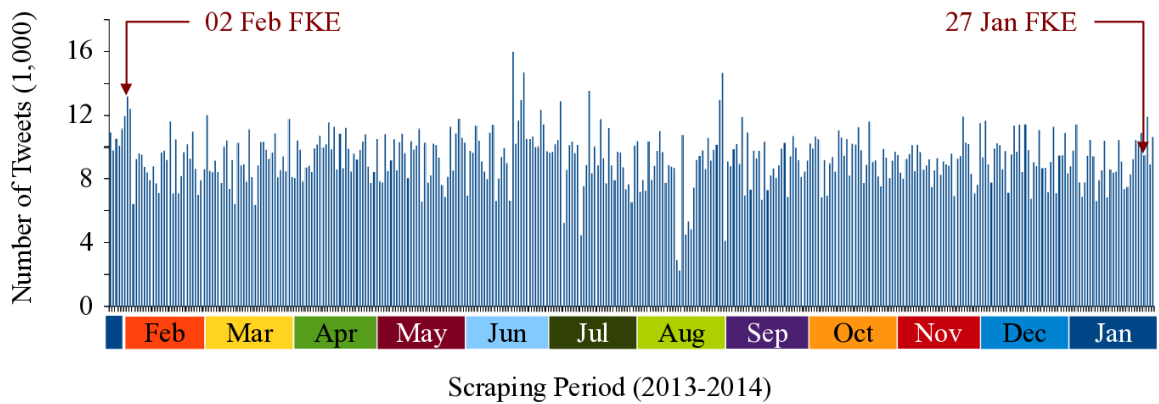, width=6in}
\caption{The daily total number of tweets collected by {\tt pScraper}.}\label{fig:4}
\end{figure}

Figure~\ref{fig:4} shows the daily number of tweets collected by {\tt pScraper} and archived to the RDBMS during the scraping period. A total of 3,423,413 tweets were gathered, of which 58\% are non-sentiment tweets (i.e., {\tt pClass} identified them as neutral tweets) and the remaining 42\% are sentiment tweets. These tweets were sent by 62,569 distinct users. On the average, each user sent 55 tweets per day. The minimum number of daily tweets is 2,269 which was sent by 1,803 unique users on 16 August 2013. The maximum number of tweets was sent on 18 June 2013 when a total of 15,995 tweets was reportedly sent by 2,804 distinct users. The average daily number of tweets is 9,278 (with $\sigma = 1,648$). 

During the collection of tweet data, it was assumed that one active account pertains to one individual only and that the sentiments of the tweets are a good proxy for the sentiment of the account users during the time of the posting of the tweet. It was also assumed that the number of pretenders and posers, although they exist, is insignificant to affect and skew the general polarity of tweets of the population. This assumption already suffices due to the absence of an official study conducted by researchers or simply by an statement from the SNS owners estimating the number of pretenders and posers. This absence is understood because the social networking site is highly dynamic in nature that estimating this number involves a lot of computational resources with little financial benefits to either the account users or SNS owners. Further, the process is already considered a gargantuan task if not an impossible one to perform. 

\subsection{Computing the CSP of Tweets via {\tt pClass}}

\begin{figure}[htb]
\centering\epsfig{file=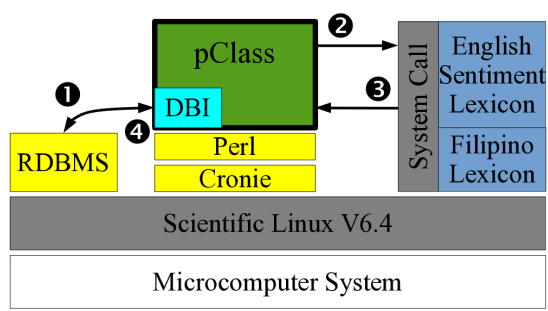, width=3in}
\caption{The conceptual relationship and data exchange among {\tt pClass}, RDBMS, Perl, Cronie, and the lexicons: (1)~{\tt pClass} extracts tweets from RDBMS via DBI; (2)~{\tt pClass} consults the lexicon via a system call; (3)~{\tt pClass} reads the sentiment from the lexicon; and (4)~{\tt pClass} updates the tweet record in the RDBMS with its estimated sentiment.}\label{fig:5}
\end{figure}

Figure~\ref{fig:5} shows the conceptual relationships among the different computer programs and {\tt pClass} in computing for the CSP of tweets. Here, {\tt pClass} extracts the tweets from the RDBMS via Perl's DBI library. Each tweet is then processed using NLP and each word is compared to the words from the English or Filipino lexicon. The lexicon is a simple look-up table which provides the possible sentiment of a given word. Once the CSP of the tweet has been estimated, {\tt pClass} then updates the tweet in the RDBMS with its computed CSP following the simple averaging technique discussed in Section II-C. 

\begin{figure}[htb]
\centering\epsfig{file=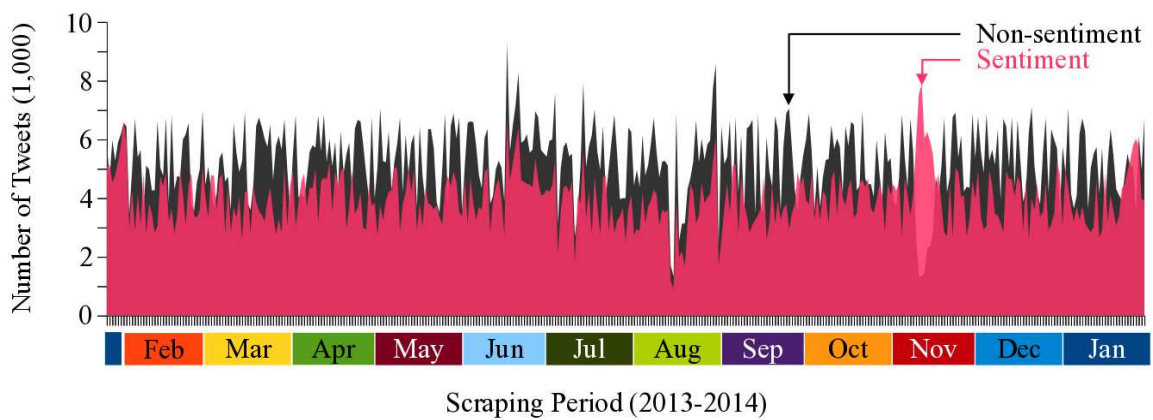, width=6in}
\caption{The daily total number of non-sentiment and sentiment tweets as computed by {\tt pClass}. Black bars are non-sentiment and red bars are sentiment tweets.}\label{fig:6}
\end{figure}

Figure~\ref{fig:6} shows the daily total number of non-sentiment (or neutral) and sentiment tweets. The average daily non-sentiment tweets is 5,288 ($\sigma = 1,238$) while the average daily sentiment tweets is 3,987 ($\sigma = 934$). On the average, the neutral tweets outnumber the sentiment tweets on a daily basis by about 1,300 except for some observed aggregate dates. In particular, there are three aggregate dates where the number of sentiment tweets is more than the number of neutral tweets: (1)~About the start of February 2013; (2)~About the second week of November 2013; and (3)~Towards the third week of January 2014. Notice that the two FKE's occur on 02 February 2013 and 16 January 2014. A 2013 event that is memorable to the minds of most Filipinos is 07-08 November 2013 when the world's strongest tropical cyclone Haiyan/Yolanda hit the Philippines. This observed pattern provides evidence that the people in Taal area have expressed their sentiments more during FKE, where the event directly affected them. The extraordinary destruction brought about by the Supertyphoon Haiyan/Yolanda has also made the people express their sentiments even though they were not directly physically affected by it.

\subsection{Temporal Analysis of Tweet Sentiments}

\begin{figure}[htb]
\centering\epsfig{file=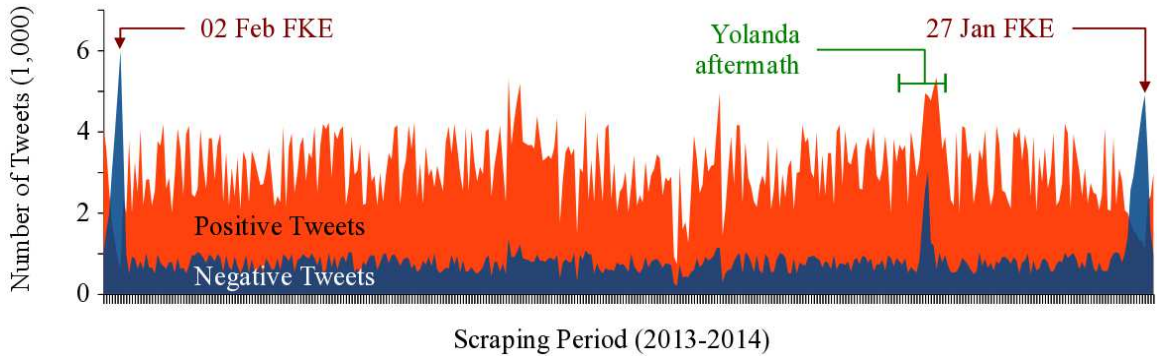, width=6in}
\caption{The daily total number of positive and negative CSP tweets. Orange bars are positive CSP tweets and blue bars are negative CSP tweets.}\label{fig:7}
\end{figure}

Figure~\ref{fig:7} shows the daily total positive and negative polarity tweets from 27 January 2013 to 30 January 2014. In general, the daily positive polarity tweets dominate the negative polarity ones except for two groups of aggregate dates: (1)~from 29 January 2015 to 03 February 2013; and (2)~from 22 to 28 January 2014. The 02 February 2013 FKE is within the first aggregate while the 16 January 2014 FKE is within the second. Equations~\ref{eqn:1} and~\ref{eqn:2} present the linear equations representing the total daily negative ($Y_\mathrm{NEG}$) and positive ($Y_\mathrm{POS}$) polarity tweets as a function of the day of the year ($X$). Both equations have their respective slopes that are not significantly different from zero. This suggests that throughout the year, both values are fairly constant.

\begin{eqnarray}
Y_\mathrm{NEG} &=&  0.16^\mathrm{ns}X +   857.13, R = 0.04\label{eqn:1}\\
Y_\mathrm{POS} &=& -0.26^\mathrm{ns}X +  3151.34, R = 0.03\label{eqn:2}
\end{eqnarray}

Even though that linear trends in Equations~\ref{eqn:1} and~\ref{eqn:2} suggest a fairly constant total daily negative and positive tweets, the respective linear trends at six days before an FKE suggest a linearly increasing negative sentiment and a linearly decreasing positive sentiment trends for both groups of aggregate dates. Equations~\ref{eqn:3} and~\ref{eqn:4} show the respective trends for the negative and positive sentiments of the 02 February 2013 FKE, while Equations~\ref{eqn:5} and~\ref{eqn:6} represent the respective trends of the negative and positive sentiments during the 16 January 2014 FKE. Without loss of generality, the independent variable $X$ in Equations~\ref{eqn:1} and~\ref{eqn:2} was replaced with $W$ in Equations~\ref{eqn:3} through~\ref{eqn:6} which now represent the reversed number of days before the observed FKE. 

\begin{eqnarray}
Y_\mathrm{NEG, 02FEB2013} &=&  915.63^\mathrm{**} W +  366.13, R =99.50\label{eqn:3}\\
Y_\mathrm{POS, 02FEB2013} &=& –579.00^\mathrm{**} W + 3809.00, R =97.01\label{eqn:4}\\
Y_\mathrm{NEG, 16JAN2014} &=&  485.49^\mathrm{**} W + 2014.80, R =99.78\label{eqn:5}\\
Y_\mathrm{POS, 16JAN2014} &=& –119.60^\mathrm{**} W + 1843.27, R =98.29\label{eqn:6}
\end{eqnarray}

\begin{figure}[htb]
\centering\epsfig{file=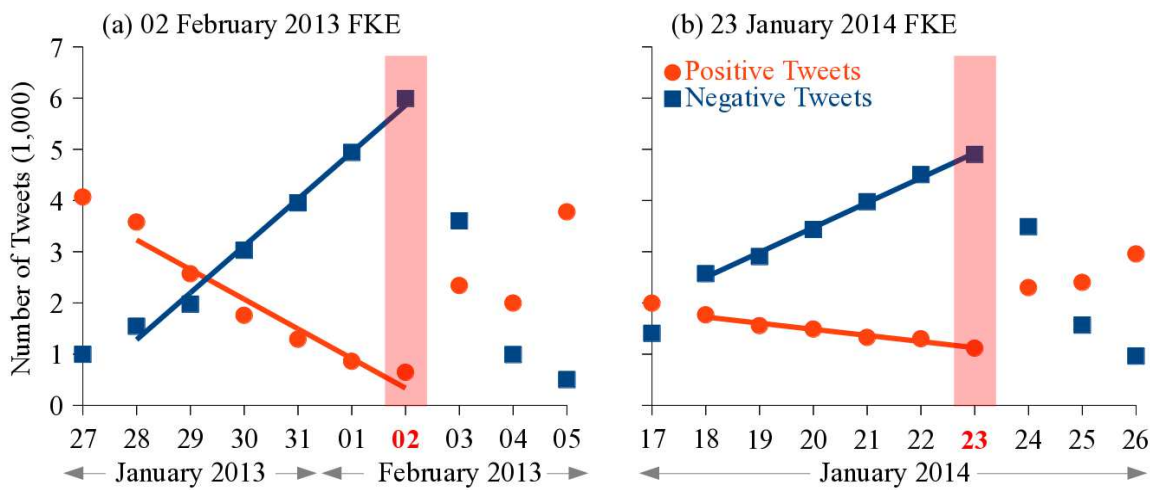, width=6in}
\caption{The respective linear trends of the negative and positive sentiment tweets around (a)~the 02 February FKE, and (b)~the 16 January 2014 FKE.}\label{fig:8}
\end{figure}

Notice here that the trends of the negative sentiments for both FKE have positive slopes suggesting increasing number of negative sentiments up to the day of the FKE (Figure~\ref{fig:8}). Concurrent to that, the trends of the positive sentiments for both FKE have negative slopes that show that the number of positive sentiments decreased significantly up to the day of the FKE.

There seems to be a pattern of resiliency exhibited in Figure~\ref{fig:8} by the trend of the number of negative and positive sentiments just merely two days after a FKE. Notice that just a day after the FKE, the number of negative sentiments decreased while the number of positive ones increased for both FKE. Two days after the FKE, the positive sentiments started to outnumber the negative ones. 

In Figure~\ref{fig:7}, other than the two aggregate dates that encompass the two FKE, there is an observable  spike in the number of negative tweets the occur from 10 to 14 November 2013. Although this spike did not outnumber the number of positive tweets within the same period, its occurrence is of interest because this pattern was observed to happen two days after Supertyphoon Haiyan/Yolanda struck the central Philippines on 07-08 November 2013. Even the typhoon did not directly physically affect the residents of Taal, they seemed to be affected  emotionally by the evidence of outpouring of the sentiment. The number of positive sentiments dominating the number of negative sentiments within these dates may show that the people in Taal were sending positive sentiments. A possible (intuitive) explanation to this observation is that the people in Taal are hopeful that those affected by the supertyphoon will ``rise up from the aftermath.'' The semantics of the tweets coming out of the textual analysis that support this intuition will be presented in another forum. 

\section{Summary and Conclusion}
This paper presented a system for sensing in real-time the sentiment of a population experiencing the onset of a FKE in Taal Lake. The FKE is seen as a rare event owing to the frequency of the occurrence of the event vis-\'a-vis the frequency of its non-occurrence. Developing a model that will provide high-fidelity prediction on when and where in the vast expanse of the lake will FKE occur using the standard regression and time series techniques have proven to be difficult because of the unbalanced nature of the dataset made complicated by the presence of missing data. Modeling techniques that rely on machine learning and predictive analytics, although successful in providing an acceptable accuracy rate, result in models that require an expensive network of sensors that must be installed not only on the surface of the vast expanse of the lake but also several meters deep along the lake's depths. Investing in such large network of sensors was foreseen to be costlier than the savings that may be obtained from avoiding the FKE. With the pervasive nature of ICT reaching even the most remote corner and used by the least expected people, it was found out from previous studies that CSP of Twitter posts can be used to track in real-time natural disasters and social phenomena. The system discussed in this work, with the aid of TAPI, accesses the texts on dated and geolocated posts on the social networking site Twitter and the corresponding CSPs of the texts are computed. Based on collected data, the significant increase in negative CSP co-occur to two FKE that separately struck Taal Lake on 02 February 2013 and on 23 January 2014. This co-occurrence of these seemingly unrelated events may give proof that CSP from Twitter may be used to augment the predictive models developed that require expensive sensing infrastructure. 


\bibliography{sentiments}
\bibliographystyle{plainnat}

\end{document}